\documentclass[10pt,conference]{IEEEtran} 
\IEEEoverridecommandlockouts
\usepackage{cite}
\usepackage{amsmath,amssymb,amsfonts}
\usepackage{algorithmic}
\usepackage{textcomp}
\usepackage{tcolorbox}
\usepackage{booktabs}
\usepackage{multirow}
\usepackage{enumitem}
\usepackage{geometry}
\usepackage{amssymb} 
\usepackage{hyperref} 
\usepackage{tabularx}
\usepackage[utf8]{inputenc}

\usepackage{amsmath}
\usepackage{tcolorbox}
\usepackage{fontawesome5}

\usepackage{graphicx}
\usepackage{array}
\usepackage{booktabs}
\usepackage{lettrine}
\usepackage{xspace}

\usepackage{csquotes}

\usepackage{xcolor}
\def\BibTeX{{\rm B\kern-.05em{\sc i\kern-.025em b}\kern-.08em
    T\kern-.1667em\lower.7ex\hbox{E}\kern-.125emX}}

\usepackage{amssymb}
\newboolean{showcomments}
\setboolean{showcomments}{true}
\newcommand{\id}[1]{$-$Id: scgPaper.tex 32478 2010-04-29 09:11:32Z oscar $-$}

\ifthenelse{\boolean{showcomments}}
{\newcommand{\nbc}[3]{
 {\colorbox{#3}{\bfseries\sffamily\scriptsize\textcolor{white}{#1}}}
 {\textcolor{#3}{\sf\small$\blacktriangleright$\emph{#2}$\blacktriangleleft$}}}
 }
{\newcommand{\nbc}[3]{}
 }

\newcommand{\ie}{\emph{i.e.},\xspace}

\newcommand{\etal}{\emph{et al.}\xspace}

\definecolor{fullyagree}{HTML}{27cc7b}
\definecolor{agree}{HTML}{83e5b5}
\definecolor{neutral}{HTML}{e5e9e7}
\definecolor{disagree}{HTML}{eeada5}
\definecolor{stronglydisagree}{HTML}{e75e4b}

\begin{document}

\title{Explaining GitHub Actions Failures with Large Language Models: Challenges, Insights, and Limitations}

\makeatletter
\newcommand{\linebreakand}{%
  \end{@IEEEauthorhalign}
  \hfill\mbox{}\par
  \mbox{}\hfill\begin{@IEEEauthorhalign}
}
\makeatother

\author{\IEEEauthorblockN{Pablo Valenzuela-Toledo\textsuperscript{1,2*}\thanks{* These authors contributed equally to the paper.}, Chuyue Wu\textsuperscript{1*}, Sandro Hernández\textsuperscript{1*}, Alexander Boll\textsuperscript{1} \\ Roman Machacek\textsuperscript{1}, Sebastiano Panichella\textsuperscript{1}, Timo Kehrer\textsuperscript{1}}
\IEEEauthorblockA{\textsuperscript{1}\emph{Software Engineering Group, University of Bern}, Bern, Switzerland\\
\textsuperscript{2}\emph{Universidad de La Frontera}, Temuco, Chile}
}


\maketitle

\begin{abstract}
GitHub Actions (GA) has become the~\emph{de facto} tool that developers use to automate software workflows, seamlessly building, testing, and deploying code. Yet when GA fails, it disrupts development, causing delays and driving up costs. Diagnosing failures becomes especially challenging because error logs are often long, complex and unstructured. 
Given these difficulties, this study explores the potential of large language models (LLMs) to generate correct, clear, concise, and actionable contextual descriptions (or summaries) for GA failures, focusing on developers' perceptions of their feasibility and usefulness.
Our results show that over 80\% of developers rated LLM explanations positively in terms of correctness for simpler/small logs. 
Overall, our findings suggest that LLMs can feasibly assist developers in understanding common GA errors, thus, potentially reducing manual analysis. 
However, we also found that improved reasoning abilities are needed to support more complex CI/CD scenarios. For instance, less experienced developers tend to be more positive on the described context, while seasoned developers prefer concise summaries. 
Overall, our work offers key insights for researchers enhancing LLM reasoning, particularly in adapting explanations to user expertise.
\end{abstract}


\section{Introduction}

GitHub Actions (GA) is an orchestration platform within GitHub that enables developers to automate tasks such as building, testing, and deploying code in integration and deployment (CI/CD) environments~\cite{golzadeh2022rise}. By minimizing manual intervention, GA streamlines development processes, boosts productivity, and is essential for maintaining efficient and stable workflows in modern software teams. Technically, GA workflows operate within specific environments, aka.\ \emph{``run/runners''}, which can be hosted on GitHub or on self-hosted servers~\cite{saroar2023developers}\cite{zhang2024developers}.

Diagnosing issues in GA workflow execution involves examining workflow execution logs~\cite{zhu2023actionsremaker}, which capture critical data—such as timestamps, success or failure indicators, error messages, and stack traces—that help identify unexpected behaviors and trace failures to their underlying causes. Additionally, environment configurations and system states provide context that supports accurate diagnosis. In today’s competitive environment, where speed and reliability are essential, quickly addressing and resolving workflow run failures is essential to limit deployment delays, increased operational costs, and compromised software stability \cite{vassallo2017tale, miller2008hundred, wilkes2023framework,zeng2024trustworthy}.

However, diagnosing GA run failures presents significant challenges due to the volume, lack of structure, and complexity of the logs~\cite{zhu2019tools, saroar2023developers}. Critical error information is often buried within extensive entries, complicating the diagnostic process, while the absence of a standardized log structure and inconsistent formats hinder automation of log analysis~\cite{saroar2023developers}. The complexity is further exacerbated by cryptic error codes, system-specific terminology, and interwoven events that obscure failure sequences, requiring specialized knowledge for accurate interpretation~\cite{chen2021experience}. Analyzing these detailed logs to uncover root causes necessitates a comprehensive understanding of both the logs and the system. 
This situation complicates pinpointing the source of failures and the actual underlying problems~\cite{alfaro2024detecting, alfaro2023mu}. Consequently, efficient filtering mechanisms are needed to isolate relevant data and facilitate the diagnosis of failures, reducing labor-intensive and error-prone tasks.

Diagnosing GA run failures demands specialized knowledge that goes beyond standard troubleshooting methods such as searching on Google or Stack Overflow. For example, developers in the \emph{bids-standard/bids-validator} repository encountered the error message, \emph{“Error: failed to load the Docker image 'bids/validator': No such file or directory”}. Resolving this error required them to analyze extensive metadata and understand both Docker configurations and the architecture of the \emph{bids-validator} tool.\footnote{\url{https://bit.ly/4fCDOGt}} Similarly, developers working in the \emph{shirasagi/shirasagi} repository processed over 20,000 lines of logs, which recorded every variable, status, and error.\footnote{\url{https://bit.ly/4fH4pSR}} These examples demonstrate how verbose GA logs often demand domain-specific expertise to interpret their context, forcing developers to tackle challenges that standard troubleshooting tools cannot address.

Recent advancements in software engineering show that large language models (LLMs) perform tasks such as code generation, summarization, understanding, and review effectively~\cite{mastropaolo2024evaluating, mastropaolo2024toward, tufano2024code, nam2024using}. We propose that developers can use LLMs to generate explanations (or summaries) that help diagnose run failures in GA workflows. LLMs recognize patterns and relationships within unstructured data, making them a promising tool for providing actionable insights to address run failures. These insights enable developers to understand and fix issues more efficiently. To our knowledge, no previous study has explored how LLMs can support the understanding of GA workflow failures.

We conducted a mixed-methods feasibility study to evaluate how LLMs explain run failures and to understand developers' perceptions of their effectiveness. We invited 811 developers to participate, and 31 accepted the invitation. This study examines four key aspects of developer perception—\emph{correctness}, \emph{conciseness}, \emph{clarity}, and \emph{actionability}—based on theoretical constructs from the literature~\cite{panichella2016impact,di2016would}. We chose these metrics because researchers have widely used them in studies on summary generation and human-based assessment, making them both relevant and well-validated.
 
Our feasibility study shows that LLMs can diagnose GA workflow failures effectively, especially in straightforward cases. Over 80\% of developers rated LLM-generated explanations as correct and clear for smaller or simpler error logs. These explanations captured essential details and provided developers with insights that helped them diagnose issues in less complex scenarios. However, LLMs struggled with intricate failures and failed to deliver the depth of reasoning required for complex CI/CD cases. Junior developers praised the contextual descriptions from LLMs, while experienced developers preferred concise explanations.

The implications of our findings are relevant primarily to developers and researchers. For developers, integrating LLMs into GA run failure diagnoses could improve troubleshooting efficiency for simpler errors by reducing the need for manual log analysis. This improvement could result in faster resolution times and greater productivity for development teams. For researchers, these findings point to new avenues for exploration, particularly in enhancing the reasoning capabilities of LLMs to address complex diagnostic tasks more effectively.

The contributions of our paper are as follows:
\begin{itemize}
    \item We conducted a \textbf{feasibility study} demonstrating LLMs' potential for interpreting GA run failures.
    \item We evaluated \textbf{prompt engineering techniques} and identified one-shot prompt tuning as the most effective approach for generating consistent and accurate GA run failure explanations.
    \item We provided \textbf{developer insights}, offering initial feedback on developers' satisfaction with LLM-generated explanations and identifying the most valued attributes of these explanations.
    \item We established a \textbf{foundation for future research}, setting the groundwork for studies on the application of LLMs in software debugging and fault localization, and encouraging further exploration and innovation in this area.
    \item We make available a \textbf{replication package} with (i) materials and datasets from our study, (ii) complete survey results, (iii) appendix with complete analysis, and  (iv) raw data to facilitate replication and support future research~\cite{valenzuela_toledo_2025_14750197}.\href{https://doi.org/10.5281/zenodo.14750197}{\includegraphics[scale=0.3]{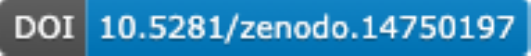}}
\end{itemize}

\section{Study Design}

The goal of this study is to evaluate the feasibility of using LLMs to explain failures in GA runs. Specifically, this evaluation examines how attributes such as \emph{correctness}, \emph{conciseness}, \emph{clarity}, and \emph{actionability} in LLM-generated explanations contribute to their diagnostic usefulness. Table~\ref{table:attributes_explanations} provides definitions for each attribute, adapted from established constructs in prior work~\cite{panichella2016impact,di2016would}. We outline three research questions that guide our study, as follows:

\textbf{RQ1}: \textit{To what extent do LLMs correctly describe the context of GitHub Action run failures according to developers?}
We investigate the developers’ perceptions of the \emph{correctness} of LLM-generated explanations in conveying the context of GA run failures. Here, \emph{correctness} indicates that the explanation is technically sound and aligns with the system and its failure's behavior.  
 
\textbf{RQ2}: \textit{To what extent do developers find generated explanations of LLMs for GitHub Action run failures clear and concise?}
We examine the clarity and conciseness of LLM-generated explanations from the developers' perspective. \emph{Clarity} refers to how understandable the explanations are, allowing developers to identify the issue quickly, while \emph{conciseness} assesses whether the explanations contain only essential information, avoiding superfluous details. These two qualities play a significant role in determining the accessibility of the explanations, as they influence how effectively developers can interpret and utilize them for diagnosing the failures and troubleshooting activities.

\textbf{RQ3}: \textit{To what extent are the descriptions of GitHub Action run failures generated by LLMs considered actionable by developers? }
We examine the \emph{actionability} of LLM-generated explanations from the developers' perspective, assessing whether these explanations provide specific and relevant information which developers can implement into a resolution for their run failures.

\begin{table}[t]
\centering
\caption{Attributes used for evaluating LLM-generated explanations of GitHub Actions run failures~\cite{panichella2016impact,di2016would}.}
\begin{tabular}{p{1.6cm}m{5.3cm}} 
\toprule
\textbf{Attribute} & \textbf{Definition} \\ 
\midrule
\emph{Correctness} & Measures the accuracy and reliability of the LLM-generated explanations in describing the actual behavior of the system, ensuring information is free from misleading content and inspires confidence in the diagnosis provided.	 \\ 
\midrule
\emph{Conciseness} & Reflects whether the explanation is efficient and avoids unnecessary information, presenting only essential details to understand and resolve the issue effectively. \\ 
\midrule
\emph{Clarity} & Assesses whether the explanation is presented in a clear and understandable manner, enabling developers to readily grasp the issue and the suggested steps. \\ 
\midrule
\emph{Actionability} & Assesses whether the explanation provides clear, step-by-step guidance that is directly implementable, enabling developers to efficiently address and resolve the failure without needing further clarification or external resources. \\ 
\bottomrule
\end{tabular}
\label{table:attributes_explanations}
\end{table}

\subsection{Data Collection}\label{sec:Survey}

We surveyed developers to investigate their perceptions on the feasibility and effectiveness of using LLMs to interpret and explain GA run failures.

\emph{Survey Design.} 

The survey consisted of 10 closed-ended statements and 2 open-ended questions as ordered in 
Fig.~\ref{fig:survey_questions}.
Statements \ref{st:one}, \ref{st:three}, \ref{st:four}, \ref{st:five}, \ref{st:six} assessed \emph{correctness} (RQ1). Statements \ref{st:two}, \ref{st:seven}, \ref{st:eight}, \ref{st:nine}, \ref{st:ten} focused on \emph{conciseness} and \emph{clarity} (RQ2). Each of these closed-ended statement used a Likert scale from 1 (strongly disagree) to 5 (strongly agree), selected based on validated assessment criteria from relevant studies in the field~\cite{barke2023grounded,cheng2024would,dakhel2023github,denny2023conversing,imai2022github,jayagopal2022exploring,jiang2022discovering}.
Finally, open-ended (OE) questions 11 and 12 addressed \emph{actionability} (RQ3).

We conducted a pilot study with experienced and novice developers to validate the survey's items, layout, and duration. Their feedback clarified the survey, confirmed a 45-60 minute completion time, and ensured clear platform guidance. The study balanced the survey’s length with its completion time by combining structured, closed-ended questions for consistency and open-ended ones for qualitative insights.

\begin{figure}[t!]
\centering
\begin{tcolorbox}[colback=gray!10, colframe=black, width=\linewidth, arc=5mm, boxrule=1pt]
\small
\textbf{Survey Statements \& Questions}
\begin{enumerate}[label=\textbf{(\arabic*)}, leftmargin=0.7cm, itemsep=0.3em]
    \item \label{st:one} The explanation accurately reflects the details and context of the GitHub Actions run failure.
    \item \label{st:two} The run failure explanation is helpful.
    \item \label{st:three} There is a low likelihood of a misleading explanation.
    \item \label{st:four} The explanation accurately diagnoses the run failure.
    \item \label{st:five} The explanation contains no inappropriate or incorrect content.
    \item \label{st:six} There is evident sound diagnostic reasoning.
    \item \label{st:seven} The explanation clearly and understandably communicates the run failure.
    \item \label{st:eight} The explanation clearly outlines the subsequent steps to take.
    \item \label{st:nine} The explanation specifically addresses my needs without being too general.
    \item \label{st:ten} I am confident in the diagnosis provided by the run failure explanation.
    \item [\textbf{$\star$(11)}] \label{st:eleven} What attributes make an error explanation valuable and effective for addressing issues related to GitHub Actions runs?
    \item [\textbf{$\star$(12)}] \label{st:twelve} Do you have any additional comments or suggestions on how we can enhance our run failure explanations?
\end{enumerate}
\end{tcolorbox}
\caption{Survey definition: Statements 1 through 10 are closed-ended, while questions 11 and 12 are open-ended (indicated with a star $\star$).}
\label{fig:survey_questions}
\end{figure}

\emph{Survey Management \& Setting.} We developed and deployed \texttt{LogExp}, a custom web tool, to effectively administer the survey and evaluate LLM-generated explanations (see Fig.~\ref{fg:LogExp}). This tool allowed for side-by-side display of both the logs (left) and their corresponding, statically deployed explanations (right), enabling participants to easily compare and assess them within a unified interface. 
In sum, the \texttt{LogExp} tool was populated with the logs and explanations of ten GA failure cases, the selection of which will be described in the sequel.

\begin{figure*}[t]
    \centering
    \includegraphics[scale=0.31]{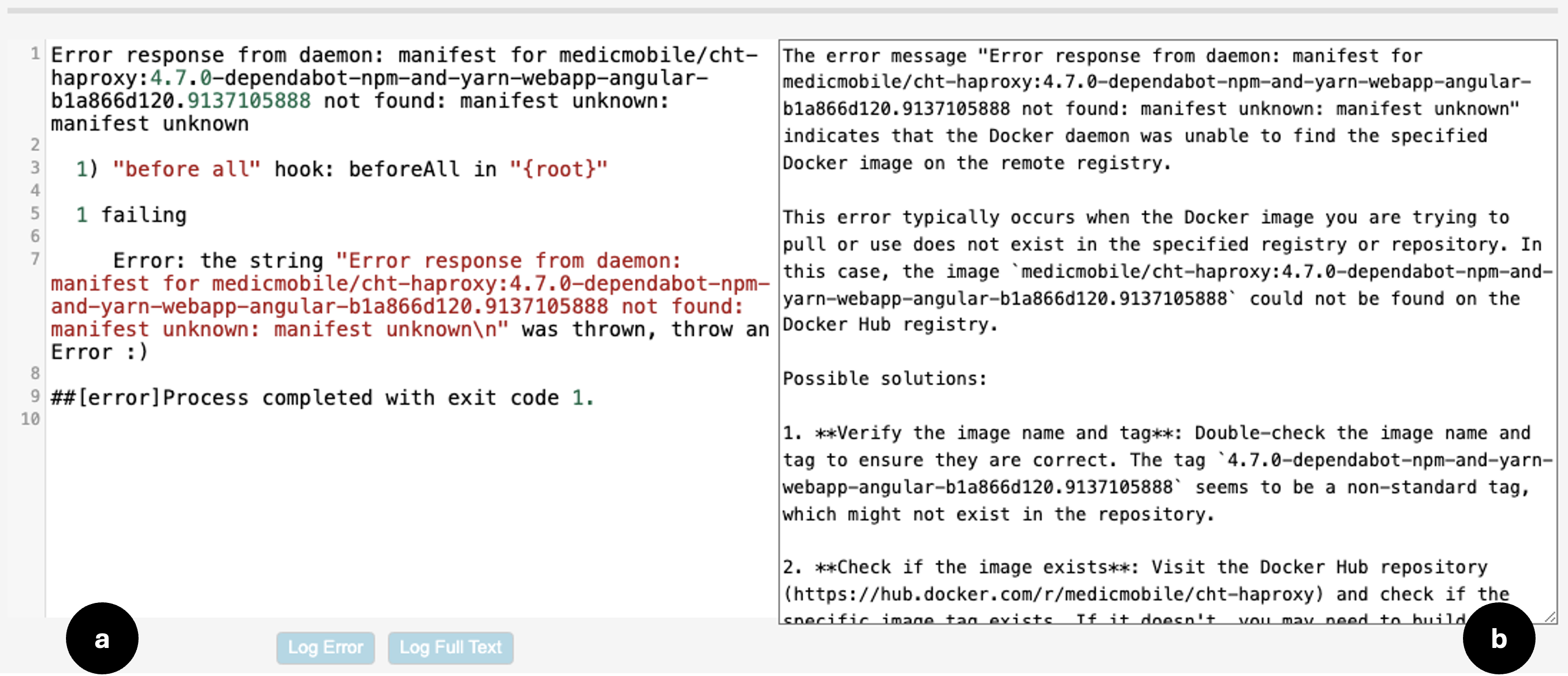}
        \caption{Partial view of the \texttt{LogExp} tool's interface. The log is displayed on the left, allowing participants to choose between viewing a summary or the full log. On the right, the corresponding textual explanation generated by the LLM is presented. Below these sections, participants encountered statements and questions specific to each case.} 
    \label{fg:LogExp}
    \vspace{-2mm}
\end{figure*}

\emph{Log Selection Procedure.}
Our study focused on GA workflow run failures in JavaScript software systems, as JavaScript ranked as the most popular programming language on GitHub in 2022~\cite{octoverse2022}. Using Dabic \etal's tool~\cite{dabic2021sampling}, we identified non-forked JavaScript repositories with at least one commit between March and May 2024 to ensure compliance with GitHub's 90-day log access limitation~\cite{github_workflow_data_2024}. We selected repositories with workflows and prioritized active, large-scale projects. To refine the dataset, we applied inclusion criteria of at least 20 contributors, 100 stars, and 20 recent commits, and we removed extreme outliers. We downloaded all failure logs and filtered out logs with fewer than 45 words. A pilot study using the knee method~\cite{weeraddana2023emse} empirically determined this threshold and ensured the logs provided enough context for actionable explanations. Logs exceeding this threshold proved more meaningful, leaving us with 348 logs. From these, we selected ten examples for our survey, which represented diverse failure cases from GA workflows, including continuous integration, deployment, and testing.

\textit{LLMs configuration}. In our pilot study, we tested combinations of LLMs and prompt techniques to integrate them into~\texttt{LogExp}. We compared three models: \texttt{Llama3} (70B), \texttt{Llama2} (70B), and \texttt{Mixtral} (8x7B). We chose \texttt{Llama3} and \texttt{Llama2} because they generated contextually accurate and structured explanations that processed CI/CD logs effectively \cite{touvron2023llama}. We included \texttt{Mixtral} for its strong performance in few-shot prompting, which reduced the need for fine-tuning and handled varied error contexts \cite{brown2020language}. We used a predefined template from the replication package [32] to generate explanations. We selected LLMs for their customizability and availability, ensuring they met our research needs. We excluded proprietary models such as GPT because their limited fine-tuning capabilities made them unsuitable for our exploratory research. Finally, we prioritized models that delivered high-quality explanations while reducing processing costs and complexity to ensure accessibility for widespread use \cite{zhang2021comparative, raffel2020exploring}.

In the pilot study, we tested three prompting techniques—\emph{zero-shot}, \emph{one-shot}, and \emph{few-shot}—on 348 GitHub Actions failure logs to evaluate their effectiveness. Prior research shows that these techniques help language models adapt to tasks with few or no examples \cite{brown2020language, radford2019language}. Five developers rated the explanations on \emph{correctness}, \emph{conciseness}, \emph{clarity}, and \emph{actionability}, and their feedback improved our prompts and models. Their evaluations identified limitations and refined our prompting techniques for each model. The pilot study showed that \texttt{Llama3} produced the most relevant and contextually accurate explanations, with \emph{one-shot} prompting providing the best balance of simplicity and accuracy.

\emph{Participants Sampling Strategy.} We used purposive sampling to select contributors familiar with GA~\cite{nagappan2013diversity,baltes2022sampling}. This method focused on developers most likely to provide relevant insights. We targeted developers who actively created, maintained, and troubleshot GA run failures. We identified these developers through their recent contributions to selected JavaScript projects on GitHub, which served as sources for our GA failure cases. To recruit participants, we contacted developers who had recently contributed to these projects. Our recruitment strategy prioritized contributors from projects with high activity to ensure their engagement in workflow development and maintenance and selected developers with experience handling and troubleshooting GA run failures.

\emph{Demographics.} We distributed the survey to 811 developers and received 31 responses, achieving a response rate of 3.82\%. Although lower than the typical range for software engineering surveys (6\% to 36\%)~\cite{smith2013improving}, this response rate aligns with exploratory studies, where 30 to 50 responses often provide sufficient insights. Most participants completed the survey in the expected time, with a median of 45 minutes. We analyzed only surveys with at least 70\% completion and retained missing data points as blanks to reduce bias.

Most respondents were male (90\%), and 10\% female. Over half (52\%) had more than 11 years of experience in software development, while 16\% had 6-10 years, 13\% had around 2 years, 10\% had 3-5 years, 6\% had 1 year or less, and 3\% reported no professional experience. In terms of education, 39\% of participants held a Master's degree in Computer Science, 6\% had a Master's in Science, Technology, Engineering, and Mathematics (STEM), and 6\% had a Master's in non-STEM fields. Bachelor’s degrees were also common, with 23\% in Computer Science or Software Engineering and 16\% in other STEM fields. A smaller portion of respondents held a high school diploma (3\%), some college or a 2-year degree (3\%), or a PhD or other advanced degrees (3\%). 

\emph{Ethics Considerations.} This study complies with the research ethics principles/regulations imposed by our university: Informed consent was obtained from all participants, and safeguards were implemented to minimize the risk of re-identification at a later time.

\subsection{Data Analysis}\label{sec:Analysis}

To analyze the data, we employed a combination of quantitative and qualitative techniques, integrating responses from closed-ended statements with insights from open-ended questions allowing for detailed, free-form responses. Our mixed-methods approach aligns with practices in prior studies (e.g.,~\cite{barke2023grounded,cheng2024would,dakhel2023github,denny2023conversing,imai2022github,jayagopal2022exploring,jiang2022discovering}) and was chosen to capture both the measurable trends and the nuanced perspectives of developers regarding LLM-generated explanations.

\emph{Quantitative Analysis.} Following best practices for survey data analysis as outlined by Kitchenham and Pfleeger \cite{kitchenham2008personal}, and Ralph \etal~\cite{ralph2020empirical}, we performed a quantitative analysis for both RQ1 and RQ2, computing descriptive statistics (mean, median, agreement). This methodological approach quantifies developers' perceptions of different aspects of LLM-generated explanations, offering a structured view of how these explanations are received. 

\emph{Qualitative Analysis.} To address RQ3, we analyzed open-ended responses from survey questions 11 and 12 using card-sorting based on Zimmermann’s three-phase methodology~\cite{zimmermann2016card}. In the preparation phase, we entered responses into an Excel sheet and split each one into cards, ensuring each card captured a single, relevant idea. During the execution phase, three authors grouped the cards independently by thematic similarity and resolved ambiguities through discussions to ensure consistency. In the analysis phase, we combined related groups into broader categories and identified key attributes that make LLM-generated explanations actionable for resolving GA run failures. We measured inter-rater agreement with Cohen’s Kappa (0.74), which showed moderate to good agreement.

\section{Correctness (RQ1)}

\begin{figure}[t] \raggedleft \includegraphics[scale=0.58 ]{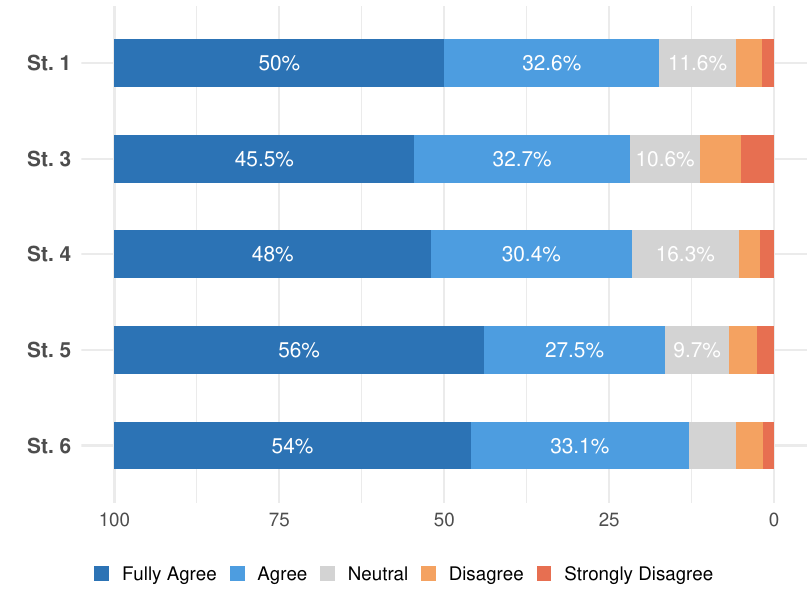} \caption{The stacked bar chart shows the levels of agreement of participants to our statements \ref{st:one}, \ref{st:three}, \ref{st:four}, \ref{st:five}, and \ref{st:six}.} \label{fig:rq1_results} \end{figure}

To address RQ1, we examined developer responses to statements \ref{st:one}, \ref{st:three}, \ref{st:four}, \ref{st:five}, and \ref{st:six}, each of which evaluated different aspects of the \emph{correctness} of LLM-generated explanations for GA run failures.

Overall, developers responded positively to these statements on correctness (see Fig.~\ref{fig:rq1_results}). St. \ref{st:one}, assessing whether explanations accurately captured the details and context of GA run failures, received 50\% full agreement and 32.6\% partial agreement, totaling 82.6\% agreement, which suggests that most developers found the explanations accurate. For St. \ref{st:three}, which focused on the likelihood of explanations being free from misleading content, 45.5\% fully agreed and 32.7\% partially agreed, resulting in an overall agreement of 78.2\%, indicating that participants generally trusted the reliability of the explanations. St. \ref{st:four}, which evaluated the precision of the explanations in diagnosing technical issues, shows 48\% full agreement and 30.4\% partial agreement, with a total of 78.4\% agreement. St. \ref{st:five}, which addressed the absence of incorrect content, showed 56\% full agreement and 27.5\% partial agreement, leading to 83.5\% agreement overall. 

Lastly, St. \ref{st:six}, which assessed the logical coherence of the explanations, received 54\% full agreement and 33.1\% partial agreement, totaling 87.1\% agreement, indicating strong approval of the explanations' logical soundness.

\begin{tcolorbox}[colback=gray!5!white, colframe=gray!75!black, title=Answer to RQ1] In summary, developers rated the \emph{correctness} of LLM-generated explanations positively, with over 80\% agreement across statements on accuracy, diagnostic precision, and logical coherence.  \end{tcolorbox}
\section{Conciseness and Clarity (RQ2)}

\begin{figure}[t]
\raggedleft
\includegraphics[scale=0.58 ]{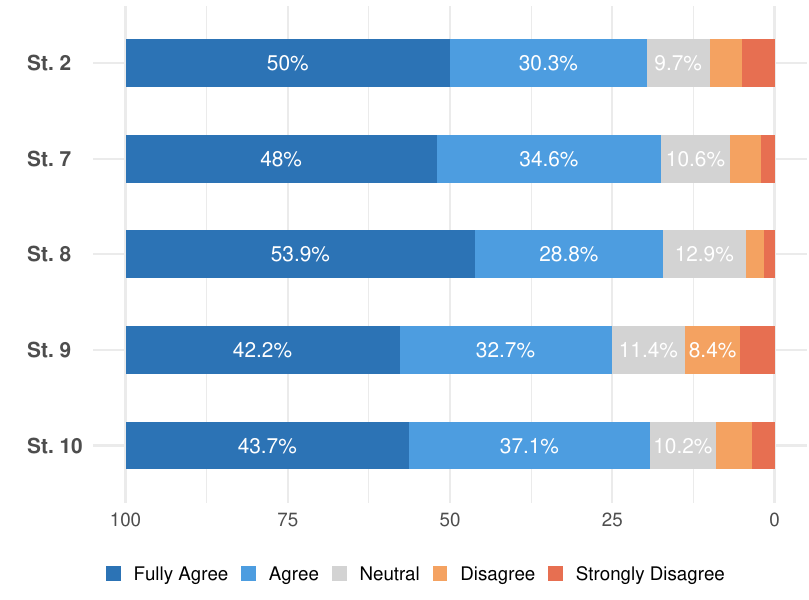}
\caption{The stacked bar chart shows the levels of agreement of participants to our statements \ref{st:two}, \ref{st:seven}, \ref{st:eight}, \ref{st:nine}, and \ref{st:ten}.} 
\label{fig:rq2_results}
\end{figure}

To address RQ2, we analyzed developer responses to statements \ref{st:two}, \ref{st:seven}, \ref{st:eight}, \ref{st:nine}, and \ref{st:ten}, each targeting distinct aspects of the conciseness and clarity of LLM-generated explanations in the context of GA run failures. 

Overall, developers responded positively to statements evaluating the conciseness and clarity of LLM-generated explanations (See Fig.~\ref{fig:rq2_results}). St. \ref{st:two}, which assessed the helpfulness of the explanations, showed 50\% full agreement and 30.3\% partial agreement, with 80.3\% agreement overall, indicating that developers found the explanations useful for diagnosing failures. St. \ref{st:seven}, directly evaluating clarity, received 48\% full agreement and 34.6\% partial agreement, reaching 82.6\% agreement overall, emphasis that the explanations were communicated clearly and understandably. St. \ref{st:eight}, focusing on outlining actionable steps, had 53.9\% full agreement and 28.8\% partial agreement, with 82.7\% agreement overall. St. \ref{st:nine}, assessing the relevance and specificity of the explanations, obtained 42.2\% full agreement and 32.7\% partial agreement, resulting in 74.9\% agreement overall, reflecting a moderate level of satisfaction with the explanations' ability to address developers' specific needs. Finally, St. \ref{st:ten}, which measured developers' confidence in the provided diagnosis, achieved 43.7\% full agreement and 37.1\% partial agreement, with 80.8\% agreement overall.

\begin{tcolorbox}[colback=gray!5!white, colframe=gray!75!black, title=  Answer to RQ2]
Developers rated the LLM-generated explanations positively for both \emph{conciseness} and \emph{clarity}. For \emph{clarity}, over 80\% of participants found the explanations easy to understand,. In terms of \emph{conciseness}, 74.5\% agreed that the explanations were specific and not overly broad. 

\end{tcolorbox}

\section{Actionability (RQ3)}

Participants evaluated the actionability of the explanations based on LLM attributes. Their responses revealed five key attributes that shape effective explanations. Each attribute reflects a distinct aspect of actionability that developers considered valuable for diagnosing and resolving GA run failures. Below, we describe these attributes and explain how they collectively enhance the actionability of LLM-generated explanations.

\subsection{Clarity of the Explanation (16\%)} This category addresses the clarity of the understanding of the explanation, avoiding excessive technical jargon and providing clear information accessible to developers at various expertise levels. Here, 
16\% of the answers highlighted this quality, indicating its importance in making explanations universally understandable. Two key attributes stand out: \emph{Clarity in error explanation} and \emph{Clarity in the steps to follow.}

\emph{Clarity in error explanation} refers to the transparency and comprehensibility of the explanation: The error is described in a straightforward manner, avoiding ambiguity. As one participant noted: 

\newcommand{\narrowquote}[1]{\vspace{3pt}\begin{quote}\setlength{\leftskip}{-6pt} \setlength{\rightskip}{-6pt}\faQuoteLeft\  \emph{#1}''\end{quote}\vspace{3pt}}

\narrowquote{I believe that a useful error explanation should get straight to the point without saying a lot of unnecessary things. Furthermore the explanation should be easily understandable by people that are just getting started so they can become better at understanding errors. Last but not least the steps to fix the issue shouldn't be too general because then a google search is better. [ID:5]}

\emph{Clarity in steps to follow} refers to providing clearly defined steps to address the failure: Each step is presented in a logical order, without unnecessary details.

\narrowquote{Clear resolution steps with examples fitting to the actual code that has one or more errors
- Clear problem explanation
- References to trustable sources. [ID:21]}

\subsection{Actionable Guidance (18\%).} This category refers to how effectively the explanation provides developers with specific, implementable steps that are directly related to resolving an specific error. 18\% of the answers highlighted the \emph{Precision in instructions} and \emph{Direct applicability of suggested steps}.

\emph{Precision in instructions} includes instructions that are specific, precise, and directly relevant to the problem: The solution steps leave no room for doubt or interpretation, allowing the developer to apply the instructions directly. As one participant noted:

\narrowquote{If the tool has access to the source code, (eg. the Actions config, project code, etc), the suggestions could actually include diffs designed to fix the issues. [ID:33]}

\emph{Direct applicability of suggested steps} describes the immediate use of the provided steps to resolve the issue: Developers can follow the instructions without needing additional research or context. 

\narrowquote{1. Cutting fluff, going straight to the point.
2. Possible steps to take to fix the problem. [ID:2]}

\subsection{Specificity of Content (18\%)} This category describes how well the explanation is tailored to the specific technical context in which the failure occurred. 18\% of the answers highlighted this quality, indicating its importance in providing relevant and accurate information for troubleshooting. The attributes in this category are: \emph{Adaptation of content to the context of the failure} and \emph{Technical accuracy in describing the problem}.

\emph{Adaptation of content to the context of the failure} refers to the altering of the explanation to its specific environment or configuration:

\narrowquote{Detecting the right context is key for error detection. If the right context is used, I also such as the examples given as long as they are small snippets. If there is more background necessary, providing a link would be my preferred way. On CI/CD topics the model is potentially useful, as you could easily try to use the recommendation and check if it resolves the issue, but this will only be helpful if the root cause of the issue is correctly identified. [ID:4]}

\emph{Technical accuracy in describing the problem} involves detailed and accurate descriptions of the error, with all technical aspects correctly presented:

\narrowquote{Narrowing down where the error happened and giving a brief explanation of what the command/process does. It helps also developers that are only 'consuming' tests to figure out faster if the failure is their fault or maybe ci/cd needs adjustments. [ID:16]}

\subsection{ Contextual Relevance (27\%)} This category addresses the inclusion of additional context or external resources to help developers understand the problem more fully. 27\% of the answers emphasized the importance of contextual relevance in error explanations. The specific attributes in this category are \emph{Inclusion of relevant links} and \emph{Explicit mention of dependencies or technical conflicts}.

\emph{Inclusion of relevant links} involves providing links to additional information about the error or its resolution:

\narrowquote{Concise information with links to relevant GitHub Actions documentation, issue trackers, or other resources that provide more in-depth information. Also, include the line numbers of the file where the issue occurs. [ID:6]}

\emph{Explicit mention of dependencies or technical conflicts} involves identifying any dependencies or conflicts that may contribute to the failure:

\narrowquote{Providing definitions of things that developers take for granted. Referencing components explicitly. Eg saying there is a dependency conflict between lib X and Lib Y. This is better than saying "there is a conflict between 2 files" You guys have nailed it. Very nice work. [ID:7]}



\subsection{Conciseness (18\%)} This category refers to the provision of brief yet informative explanations that enable developers to understand the problem efficiently, focusing on essential information omitting unnecessary details. 18\% of the answers highlighted the importance of conciseness in explanations. The specific attributes in this category are the \emph{Ability to quickly summarize the cause of the failure} and the \emph{Concise presentation of resolution steps.}

The \emph{ability to quickly summarize the cause of the failure} involves identifying the root cause of the issue:

\narrowquote{There are too many details in the procedure, the sentences such as "Edit the file" and "Commit your changes" are superfluous for a developer. [ID:34]}

\emph{Concise presentation of resolution steps} includes only the necessary actions to resolve the issue, presented clearly and directly. As another participant shared:

\narrowquote{The LLM generated texts are a little bit of being too long, it could be briefer without losing content. [ID:27]}


\begin{tcolorbox}[colback=gray!5!white, colframe=gray!75!black, title=Answer to RQ3]
Effective explanations for GitHub Actions run failures include five key attributes: \emph{clarity}, which provides straightforward information; \emph{actionable guidance}, offering precise steps for resolution; \emph{specificity}, adapting explanations to the technical context; \emph{contextual relevance}, adding links or details about dependencies; and \emph{conciseness}, ensuring only essential information is presented.
\end{tcolorbox}

\section{Discussion and Implications}

This section summarizes key findings for each research question, discussing relevant confounding factors and implications for future research and practical applications. In addition to the results we systematically presented in the last section, we further analyzed various metrics that we present in this section.

\textbf{RQ1: Correctness of LLM-Generated Explanations}. Our results indicate that LLMs provide accurate explanations for simpler failures, especially with structured and concice logs. This aligns with previous studies that stress the role of well-organized logs in facilitating error diagnosis in CI/CD environments. For instance, Vassallo~\etal highlight that structured data in CI logs improves diagnostic efficiency, with LLMs benefiting from reduced ambiguity and focused log information~\cite{vassallo2017tale}. 

LLMs struggle with verbose and unstructured logs that obscure key information. Sallou~\etal highlight LLMs' inconsistent performance on unstructured data, impacting their accuracy in real-world scenarios~\cite{sallou2024breaking}. In our work, excessive detail in log entries hid critical error messages, revealing the need for preprocessing to extract relevant information. For instance, one participant noted, \emph{``... providing definitions of things that developers take for granted.
- Referencing components explicitly. Eg saying there is a dependency conflict between lib X and Lib Y''.} would greatly enhance correctness. This suggests that developers cannot use LLMs and prompts directly; they must tailor preprocessing for specific contexts. Additionally, preprocessing is crucial for large logs due to LLMs' token limits~\cite{da2024chatgpt}.

Our analysis showed that 80\% of participants used CI/CD in their primary activities. This widespread use shaped their expectations for error explanations, as CI/CD users gave higher median scores of 4.5 compared to 3.2 from participants who did not rely on CI/CD as a primary activity. CI/CD users also rated explanations more consistently, with a standard deviation of 0.5, while non-users had a standard deviation of 1.2. These findings suggest that familiarity with CI/CD practices unifies perceptions of error explanations. Prior research \cite{xia2023automated,vassallo2017tale} supports this observation, showing that structured input improves diagnostic efficiency in CI/CD contexts.

Our analysis of log length revealed that shorter logs received higher and more consistent ratings, while longer logs produced greater variability in responses. This result suggests that log length influences how participants perceive the explanations. These findings align with prior research showing the value of adaptive summaries tailored to user experience levels \cite{panichella2016impact}. Customizing explanations based on developers' expertise and accounting for log-specific contextual factors could improve correctness.

\textbf{RQ2: Conciseness and Clarity of LLM-Generated Explanations}.

The study finds that LLMs generally provide clear and concise explanations for simple GA run failures, particularly when logs are structured and free from excessive complexity or irrelevant information. This observation supports the notion that conciseness is key to effective communication in technical contexts. Previous research by Vassallo \etal \cite{vassallo2017tale} emphasizes that well-organized logs facilitate error diagnosis in CI/CD environments, thereby enhancing clarity. Furthermore, Xia \etal \cite{xia2023automated} suggest that LLMs leverage structured input to produce outputs that are not only reliable but also easy to understand.

We observed no significant differences in conciseness or clarity based on log complexity (\ie cryptic error codes, system-specific terminology, and interwoven events that obscure the failure sequence). However, participants occasionally noted redundancy in the explanations. This aligns with the validity threats discussed by Sallou \etal, who warn that LLMs may introduce unnecessary details that obscure the main issue \cite{sallou2024breaking}. Similarly, Chen \etal highlight that verbosity in LLM-generated explanations can reduce their effectiveness when excessive information overshadows important details \cite{chen2023chatgpt}. One participant remarked, \emph{``narrowing down where the error happened and giving a brief explanation of what the command/process does helps developers figure out faster if the failure is their fault or maybe CI/CD needs adjustments.'' [ID:16]}.

\textbf{RQ3: Actionability of LLM-Generated Explanations}. 
The study identifies five key dimensions—clarity, actionable guidance, specificity, contextual relevance, and conciseness—as central to the actionability of LLM-generated explanations for GA failures, independent of error complexity or developer experience. This aligns with findings by Xia~\etal, who emphasize that clarity and relevance are foundational for effective automated diagnostics in software engineering \cite{xia2023automated}. Additionally, Chen~\etal note that without focused guidance, LLM explanations risk becoming too generalized, which can reduce their practical impact for both novice and experienced users \cite{chen2023chatgpt}. Our results suggest that focusing on these dimensions may benefit all experience levels.

Interestingly, the study found minimal variation in actionability perceptions across experience levels, suggesting that LLMs can support a wide range of users with consistent guidance. However, Sallou~\etal. discuss the limitations of static LLM outputs, especially in failing to provide the in-depth insights that advanced users might seek \cite{sallou2024breaking}. This suggests a potential improvement area: developing adaptive models that could provide variable explanation depth based on user experience, thereby addressing both basic and advanced informational needs.

Future research could explore \textit{dynamic adaptation} in LLM explanations, as proposed by Ye~\etal, who argue that personalized guidance based on familiarity enhances the value of automated explanations \cite{ye2023assessing}. Such adaptability could ensure that LLMs provide universally useful guidance and cater to the specific cognitive demands of developers in complex CI/CD environments.

Furthermore, quality of generated output can be improved using numerous ways. Firstly, \textit{consensus based generation}~\cite{bakerconsensus} would allow different models to unite their response and utilize different views. Secondly, models have shown improved reasoning and generation with chain-of-thought prompting~\cite{weiprompting}, for further explanation and improvement of results. Lastly, instruction prompting~\cite{benfengexpertprompt} can be utilized for more fine-grained control over generated text, for instance output generation based on the experience of the developer. 

\textbf{Implications for Researchers.}
Our results suggest several research directions for improving LLMs’ ability to support diagnostics in CI/CD environments. For instance, researchers could focus on refining LLMs to better handle diverse log structures by developing preprocessing techniques to filter essential information within verbose or unstructured logs. An example of this can be seen in the GA log from the BIDS Validator project, which shows how clarity can be enhanced by emphasizing key details.\footnote{\url{https://bit.ly/4fCDOGt}} Such preprocessing could enable LLMs to generate accurate explanations even in complex log contexts. Additionally, adapting LLM outputs to user-specific contexts based on expertise level could enhance usability, providing concise, high-level insights for advanced users and step-by-step guidance for beginners. Investigating these adaptive mechanisms and fine-tuning LLM responses for both novice and expert users offers a promising approach for making LLM-powered diagnostics more universally applicable.

\textbf{Implications for Developers.} For developers, the findings suggest practical improvements for LLM-powered diagnostic tools that could increase explanation accuracy and relevance. Specifically, integrating log preprocessing features to structure data before analysis can help LLMs focus on the most critical information, resulting in clearer and more actionable explanations. Additionally, adjustable explanation levels could be defined by variables such as the developer's experience level, the complexity of the run failure, and the context of the log data. For instance, developers could toggle between concise summaries for routine errors and comprehensive guidance for more complex issues. Such flexibility would make LLM tools more adaptable to real-world CI/CD workflows, improving troubleshooting efficiency by delivering the appropriate level of detail tailored to each developer’s needs.

\textbf{Practical Applications.} The findings from RQ1, RQ2, and RQ3 suggest that LLM-powered explanations are generally perceived as accurate, concise, clear, and actionable by both novice and expert users, with the potential to enhance CI/CD troubleshooting efficiency by reducing manual intervention. For instance, a GitHub Community discussion describes persistent run failure in GA due to frequent misconfigurations in permissions and authentication settings, resulting in a  run failure\footnote{\url{https://bit.ly/3AoGKYr}}. Developers in the thread discuss solutions such as adjusting permissions and optimizing job conditions to prevent these failures. This example shows how LLMs could provide automated diagnostics by identifying common misconfigurations and suggesting corrective actions for recurring issues. By offering context-sensitive insights, LLMs could help reduce the high rate of failures in CI/CD workflows, streamlining the troubleshooting process and improving workflow reliability.

\section{Threats to Validity} 

In this section, we outline possible threats to the validity of our study and show how we mitigated them.

\textit{Construct Validity}. 
Due to the fact that our study was performed in a remote setting in which
participants could work on the tasks at their own discretion, we could not oversee their behavior. However, we anticipate the procedure of the experiments and all relevant details on how to conduct the experiments with a survey guiding the various steps of participants.
 
In addition, our study limited the assessment of the log summaries based on four main metrics (i.e., Correctness, Conciseness, Clarity, and Actionability), which may theoretically restrict the generalization of our findings to these specific metrics.
However, we selected them because they are widely adopted in academia on previous summary generation and human-based assessment studies~\cite{panichella2016impact,di2016would}.

\textit{Internal Validity}. 
As LLMs generate their responses in a probabilistic manner, additional responses to the same prompts could have been perceived as better or worse than those from our survey. A setup where participants rate multiple explanations to the same logs and prompts could have controlled for this. However, our study-setup was already time-consuming with 45 minutes, and we wanted to ensure enough participants would complete our survey.
Moreover, the study participants rated the log summaries based on their perceived quality of description generated by the approach.
To limit the risks of biased assessments, we clarified the meaning of the criteria used to assess the log summaries before the experiments to the participants. Moreover, we specifically targeted developers who are actively involved in creating, maintaining, and troubleshooting GA run failures. 
Finally, the varying complexity of logs may lead to varying perceptions of the corresponding generated summaries. Rather than a threat this is more of a potential confounding factor we actually investigate in the context of our study.

\textit{Threats to External Validity}. 
The selection of participants does not represent the general software developer population. Instead, we specifically targeted relevant developers (i.e., actively involved in creating, maintaining, and troubleshooting GA run failures), primarily by applying a sampling approach that selects a representative set of contributors familiar with GA~\cite{nagappan2013diversity,baltes2022sampling}.
To address other potential bias, we ensured diversity in terms of developers experience with GA, reducing the influence of factors beyond professional background.
Furthermore, from our results one cannot predict the quality of explanations of other LLMs and different prompting techniques, which we did not test systematically. Our study does not aim to optimize LLM performance or determine the best model or prompting strategy. Instead, it explores how LLMs, as an emerging technology, can enhance developers' understanding of GitHub Actions run failures through actionable explanations. This exploratory work addresses a gap in CI/CD troubleshooting by identifying trends and providing foundational insights rather than exhaustive comparisons.
. However, our internal pilot study showed that the three LLMs \texttt{Llama3}, \texttt{Llama2}, and \texttt{Mixtral} with various prompting techniques returned explanations of similar quality. 
Recent studies suggest that modern LLMs achieve hight quality in summaries and explanations in specific contexts \cite{liu2023responsible}.
Finally, further studies will be required to see whether results generalize to other logs of other programming languages, an effort which we leave for future work.
\section{Related Work}

This section reviews advancements in automatic summarization within software engineering.

\emph{(i) Build failure detection and resolution tools.} Several studies address build failure detection in CI/CD environments. Vassallo \etal introduced BART to summarize and resolve Maven build failures and conducted a broad analysis identifying common CI failure causes across open-source projects, providing classification methods for typical errors~\cite{vassallo2020every, vassallo2017tale}. Rausch \etal also examined CI failures, focusing on error categories and the importance of build stability to support developer workflows~\cite{rausch2017empirical}. Additionally, Alfaro \etal proposed Microprints, a visualization tool to identify errors in CI/CD logs, although it does not specifically address run failure explanations~\cite{alfaro2024detecting, alfaro2023mu}.

\emph{(ii) Challenges and developments in source code summarization and metric evaluation.} The lack of standardized datasets complicates source code summarization, as LeClair \etal highlight the challenges inconsistent data presents for research outcomes~\cite{LeClair2019Dataset}. Moreno \etal advocate for improved benchmarks to maximize the utility of automatic summarization for stakeholders~\cite{Moreno2018SoftwareSummarization}. Additionally, Tarar and Zhang’s work on bug report summarization demonstrates efficiency gains by leveraging syntactic and semantic similarities~\cite{Tarar2019BugSummarization, Zhang2020Rencos}. Traditional metrics such as BLEU and ROUGE are critiqued by Stapleton \etal and Roy \etal for inadequacies in reflecting developer needs, while newer metrics such as SIDE, which employ contrastive learning, provide better alignment with human evaluations~\cite{Stapleton2020ComprehensionStudy, Roy2021ReassessingMetrics, Banerjee2005METEOR, Haque2022SemanticSimilarity, Mastropaolo2024SIDE}.

\emph{(iii) The application of neural and federated learning models in summarization.} Neural and federated learning models have significantly advanced summarization techniques. Iyer \etal's CODE-NN, a neural attention model, surpassed previous methods for summarizing C\# and SQL code~\cite{Iyer2016NeuralSummarization}. Kumar \etal proposed FedLLM, a federated learning approach to code summarization that ensures data privacy while maintaining centralized model performance~\cite{Kumar2024FedLLM}. Structured summaries developed by Moreno \etal and by McBurney and McMillan aid developers in understanding Java code by using stereotypes and call graphs~\cite{Moreno2013JavaSummarization, McBurney2016JavaContextSummarization}. Other works, such as those by Rastkar and Haiduc, have improved bug report handling and code entity summarization~\cite{Rastkar2014BugSummarization, Haiduc2010TextSummarization}. Dabrowski’s review of app review analysis emphasizes their growing importance in software development, and Nazar and Panichella show the utility of automated summarization in software maintenance and debugging~\cite{Dabrowski2022AppReview, Nazar2016SummarizingSoftwareArtifacts, Panichella2018SummarizationTechniques}.

Collectively, these studies indicate a need for diagnostic tools that not only identify errors but also provide actionable explanations for complex, unstructured log data. While previous work has advanced detection, categorization, and summarization, interpreting run failures in CI/CD workflows such as those in GA remains a challenge. Our research builds on these foundations, aiming to address this gap by leveraging large language models to enhance troubleshooting support.

\section{Conclusion}

Our findings show that developers appreciate clear, concise LLM explanations for simpler issues. However, as error complexity grows, LLMs often lack accuracy and detail, underscoring the need for improvement in complex troubleshooting. Future work should focus on enhancing LLMs' capacity to contextualize complex errors in GA workflows and tailoring explanations for different expertise levels. Integrating log preprocessing and customizable explanation settings can potentially further improve usability, helping LLMs prioritize critical information and meet diverse troubleshooting needs. Advancing in these areas may lead to more effective diagnostic tools, boosting developer productivity and CI/CD stability.

\textbf{Acknowledgments}. The authors would like to thank the
Swiss Group Software Engineering (CHOOSE) for sponsoring the trip to the conference.

\bibliography{references.bib}

\begin{thebibliography}{10}
\providecommand{\url}[1]{#1}
\csname url@samestyle\endcsname
\providecommand{\newblock}{\relax}
\providecommand{\bibinfo}[2]{#2}
\providecommand{\BIBentrySTDinterwordspacing}{\spaceskip=0pt\relax}
\providecommand{\BIBentryALTinterwordstretchfactor}{4}
\providecommand{\BIBentryALTinterwordspacing}{\spaceskip=\fontdimen2\font plus
\BIBentryALTinterwordstretchfactor\fontdimen3\font minus \fontdimen4\font\relax}
\providecommand{\BIBforeignlanguage}[2]{{%
\expandafter\ifx\csname l@#1\endcsname\relax
\typeout{** WARNING: IEEEtran.bst: No hyphenation pattern has been}%
\typeout{** loaded for the language `#1'. Using the pattern for}%
\typeout{** the default language instead.}%
\else
\language=\csname l@#1\endcsname
\fi
#2}}
\providecommand{\BIBdecl}{\relax}
\BIBdecl

\bibitem{golzadeh2022rise}
M.~Golzadeh, A.~Decan, and T.~Mens, ``On the rise and fall of {CI} services in {GitHub},'' in \emph{2022 IEEE International Conference on Software Analysis, Evolution and Reengineering}.\hskip 1em plus 0.5em minus 0.4em\relax IEEE, 2022, pp. 662--672.

\bibitem{saroar2023developers}
S.~G. Saroar and M.~Nayebi, ``Developers’ perception of github actions: A survey analysis,'' in \emph{Proceedings of the 27th International Conference on Evaluation and Assessment in Software Engineering}, 2023, pp. 121--130.

\bibitem{zhang2024developers}
Y.~Zhang, Y.~Wu, T.~Chen, T.~Wang, H.~Liu, and H.~Wang, ``How do developers talk about {GitHub} actions? evidence from online software development community,'' in \emph{Proceedings of the 46th IEEE/ACM International Conference on Software Engineering}, 2024, pp. 1--13.

\bibitem{zhu2023actionsremaker}
H.-N. Zhu, K.~Z. Guan, R.~M. Furth, and C.~Rubio-Gonzalez, ``Actionsremaker: Reproducing github actions,'' in \emph{2023 IEEE/ACM 45th International Conference on Software Engineering: Companion Proceedings}.\hskip 1em plus 0.5em minus 0.4em\relax IEEE, 2023, pp. 11--15.

\bibitem{vassallo2017tale}
C.~Vassallo, G.~Schermann, F.~Zampetti, D.~Romano, P.~Leitner, A.~Zaidman, M.~Di~Penta, and S.~Panichella, ``A tale of ci build failures: An open source and a financial organization perspective,'' in \emph{2017 IEEE international conference on software maintenance and evolution}.\hskip 1em plus 0.5em minus 0.4em\relax IEEE, 2017, pp. 183--193.

\bibitem{miller2008hundred}
A.~Miller, ``A hundred days of continuous integration,'' in \emph{Agile 2008 conference}.\hskip 1em plus 0.5em minus 0.4em\relax IEEE, 2008, pp. 289--293.

\bibitem{wilkes2023framework}
B.~Wilkes, A.~M.~P. Milani, and M.-A. Storey, ``A framework for automating the measurement of devops research and assessment (dora) metrics,'' in \emph{2023 IEEE International Conference on Software Maintenance and Evolution}.\hskip 1em plus 0.5em minus 0.4em\relax IEEE, 2023, pp. 62--72.

\bibitem{zeng2024trustworthy}
Z.~Zeng, T.~Xiao, M.~Lamothe, H.~Hata, and S.~McIntosh, ``How trustworthy is your ci accelerator? a comparison of the trustworthiness of ci acceleration products,'' \emph{IEEE Software}, 2024.

\bibitem{zhu2019tools}
J.~Zhu, S.~He, J.~Liu, P.~He, Q.~Xie, Z.~Zheng, and M.~R. Lyu, ``Tools and benchmarks for automated log parsing,'' in \emph{2019 IEEE/ACM 41st International Conference on Software Engineering: Software Engineering in Practice}.\hskip 1em plus 0.5em minus 0.4em\relax IEEE, 2019, pp. 121--130.

\bibitem{chen2021experience}
Z.~Chen, J.~Liu, W.~Gu, Y.~Su, and M.~R. Lyu, ``Experience report: Deep learning-based system log analysis for anomaly detection,'' \emph{arXiv preprint arXiv:2107.05908}, 2021.

\bibitem{alfaro2024detecting}
S.~Alfaro, A.~Bergel, and J.~Simmonds, ``Detecting ci/cd workflow errors through visual inspection of logs,'' \emph{Authorea Preprints}, 2024.

\bibitem{alfaro2023mu}
------, ``mu printgen: Supporting workflow logs analysis through visual microprint,'' in \emph{2023 IEEE Working Conference on Software Visualization}.\hskip 1em plus 0.5em minus 0.4em\relax IEEE, 2023, pp. 45--49.

\bibitem{mastropaolo2024evaluating}
A.~Mastropaolo, M.~Ciniselli, M.~Di~Penta, and G.~Bavota, ``Evaluating code summarization techniques: A new metric and an empirical characterization,'' in \emph{Proceedings of the IEEE/ACM 46th International Conference on Software Engineering}, 2024, pp. 1--13.

\bibitem{mastropaolo2024toward}
A.~Mastropaolo, F.~Zampetti, G.~Bavota, and M.~Di~Penta, ``Toward automatically completing {GitHub} workflows,'' in \emph{Proceedings of the 46th IEEE/ACM International Conference on Software Engineering}, 2024, pp. 1--12.

\bibitem{tufano2024code}
R.~Tufano, O.~Dabi{\'c}, A.~Mastropaolo, M.~Ciniselli, and G.~Bavota, ``Code review automation: strengths and weaknesses of the state of the art,'' \emph{IEEE Transactions on Software Engineering}, 2024.

\bibitem{nam2024using}
D.~Nam, A.~Macvean, V.~Hellendoorn, B.~Vasilescu, and B.~Myers, ``Using an llm to help with code understanding,'' in \emph{Proceedings of the IEEE/ACM 46th International Conference on Software Engineering}, 2024, pp. 1--13.

\bibitem{panichella2016impact}
S.~Panichella, A.~Panichella, M.~Beller, A.~Zaidman, and H.~C. Gall, ``The impact of test case summaries on bug fixing performance: An empirical investigation,'' in \emph{Proceedings of the 38th international conference on software engineering}, 2016, pp. 547--558.

\bibitem{di2016would}
A.~Di~Sorbo, S.~Panichella, C.~V. Alexandru, J.~Shimagaki, C.~A. Visaggio, G.~Canfora, and H.~C. Gall, ``What would users change in my app? summarizing app reviews for recommending software changes,'' in \emph{Proceedings of the 2016 24th ACM SIGSOFT international symposium on foundations of software engineering}, 2016, pp. 499--510.

\bibitem{valenzuela_toledo_2025_14750197}
\BIBentryALTinterwordspacing
P.~Valenzuela-Toledo, C.~Wu, S.~Hernández, A.~Boll, R.~Machacek, S.~Panichella, and T.~Kehrer, ``Explaining github actions failures with large language models: Challenges, insights, and limitations,'' Jan. 2025. [Online]. Available: \url{https://doi.org/10.5281/zenodo.14750197}
\BIBentrySTDinterwordspacing

\bibitem{barke2023grounded}
S.~Barke, M.~B. James, and N.~Polikarpova, ``Grounded copilot: How programmers interact with code-generating models,'' \emph{Proceedings of the ACM on Programming Languages}, vol.~7, no. OOPSLA1, pp. 85--111, 2023.

\bibitem{cheng2024would}
R.~Cheng, R.~Wang, T.~Zimmermann, and D.~Ford, ``“it would work for me too”: How online communities shape software developers’ trust in ai-powered code generation tools,'' \emph{ACM Transactions on Interactive Intelligent Systems}, vol.~14, no.~2, pp. 1--39, 2024.

\bibitem{dakhel2023github}
A.~M. Dakhel, V.~Majdinasab, A.~Nikanjam, F.~Khomh, M.~C. Desmarais, and Z.~M.~J. Jiang, ``Github copilot ai pair programmer: Asset or liability?'' \emph{Journal of Systems and Software}, vol. 203, p. 111734, 2023.

\bibitem{denny2023conversing}
P.~Denny, V.~Kumar, and N.~Giacaman, ``Conversing with copilot: Exploring prompt engineering for solving cs1 problems using natural language,'' in \emph{Proceedings of the 54th ACM Technical Symposium on Computer Science Education V. 1}, 2023, pp. 1136--1142.

\bibitem{imai2022github}
S.~Imai, ``Is github copilot a substitute for human pair-programming? an empirical study,'' in \emph{Proceedings of the ACM/IEEE 44th International Conference on Software Engineering: Companion Proceedings}, 2022, pp. 319--321.

\bibitem{jayagopal2022exploring}
D.~Jayagopal, J.~Lubin, and S.~E. Chasins, ``Exploring the learnability of program synthesizers by novice programmers,'' in \emph{Proceedings of the 35th Annual ACM Symposium on User Interface Software and Technology}, 2022, pp. 1--15.

\bibitem{jiang2022discovering}
E.~Jiang, E.~Toh, A.~Molina, K.~Olson, C.~Kayacik, A.~Donsbach, C.~J. Cai, and M.~Terry, ``Discovering the syntax and strategies of natural language programming with generative language models,'' in \emph{Proceedings of the 2022 CHI Conference on Human Factors in Computing Systems}, 2022, pp. 1--19.

\bibitem{octoverse2022}
\BIBentryALTinterwordspacing
GitHub, ``Top programming languages of 2022 - github octoverse,'' 2022, accessed: 2024-11-07. [Online]. Available: \url{https://octoverse.github.com/2022/top-programming-languages}
\BIBentrySTDinterwordspacing

\bibitem{dabic2021sampling}
O.~Dabic, E.~Aghajani, and G.~Bavota, ``Sampling projects in github for msr studies,'' in \emph{2021 IEEE/ACM 18th International Conference on Mining Software Repositories}.\hskip 1em plus 0.5em minus 0.4em\relax IEEE, 2021, pp. 560--564.

\bibitem{github_workflow_data_2024}
{GitHub, Inc.}, ``Storing and sharing data from a workflow,'' \url{https://docs.github.com/en/actions/writing-workflows/choosing-what-your-workflow-does/storing-and-sharing-data-from-a-workflow}, 2024, accessed: 2024-08-26.

\bibitem{weeraddana2023emse}
N.~R. Weeraddana, X.~Xu, M.~Alfadel, S.~McIntosh, and M.~Nagappan, ``{An Empirical Comparison of Ethnic and Gender Diversity of DevOps and non-DevOps Contributions to Open-Source Projects},'' \emph{Empirical Software Engineering}, vol.~28, no. 150, p. 1–37, 2023.

\bibitem{touvron2023llama}
H.~Touvron, T.~Lavril, G.~Izacard, X.~Martinet, M.-A. Lachaux, T.~Lacroix \emph{et~al.}, ``Llama: Open and efficient foundation language models,'' in \emph{arXiv preprint arXiv:2302.13971}, 2023.

\bibitem{brown2020language}
T.~B. Brown, B.~Mann, N.~Ryder, M.~Subbiah, J.~Kaplan, P.~Dhariwal, A.~Neelakantan, P.~Shyam, G.~Sastry, A.~Askell \emph{et~al.}, ``Language models are few-shot learners,'' in \emph{Advances in Neural Information Processing Systems}, vol.~33, 2020, pp. 1877--1901.

\bibitem{zhang2021comparative}
H.~Zhang, Y.~Sun, and Y.~Qi, ``Comparative analysis of pre-trained language models for natural language understanding,'' in \emph{Proceedings of the 2021 Conference on Empirical Methods in Natural Language Processing}, 2021.

\bibitem{raffel2020exploring}
C.~Raffel, N.~Shazeer, A.~Roberts, K.~Lee, S.~Narang, M.~Matena, Y.~Zhou, W.~Li, and P.~J. Liu, ``Exploring the limits of transfer learning with a unified text-to-text transformer,'' \emph{Journal of Machine Learning Research}, vol.~21, no. 140, pp. 1--67, 2020.

\bibitem{radford2019language}
A.~Radford, J.~Wu, R.~Child, D.~Luan, D.~Amodei, and I.~Sutskever, ``Language models are unsupervised multitask learners,'' in \emph{OpenAI Technical Report}, 2019.

\bibitem{nagappan2013diversity}
M.~Nagappan, T.~Zimmermann, and C.~Bird, ``Diversity in software engineering research,'' in \emph{Proceedings of the 2013 9th joint meeting on foundations of software engineering}, 2013, pp. 466--476.

\bibitem{baltes2022sampling}
S.~Baltes and P.~Ralph, ``Sampling in software engineering research: A critical review and guidelines,'' \emph{Empirical Software Engineering}, vol.~27, no.~4, p.~94, 2022.

\bibitem{smith2013improving}
E.~Smith, R.~Loftin, E.~Murphy-Hill, C.~Bird, and T.~Zimmermann, ``Improving developer participation rates in surveys,'' in \emph{2013 6th International workshop on cooperative and human aspects of software engineering}.\hskip 1em plus 0.5em minus 0.4em\relax IEEE, 2013, pp. 89--92.

\bibitem{kitchenham2008personal}
B.~A. Kitchenham and S.~L. Pfleeger, ``Personal opinion surveys,'' in \emph{Guide to advanced empirical software engineering}.\hskip 1em plus 0.5em minus 0.4em\relax Springer, 2008, pp. 63--92.

\bibitem{ralph2020empirical}
P.~Ralph, N.~b. Ali, S.~Baltes, D.~Bianculli, J.~Diaz, Y.~Dittrich, N.~Ernst, M.~Felderer, R.~Feldt, A.~Filieri \emph{et~al.}, ``Empirical standards for software engineering research,'' \emph{arXiv preprint arXiv:2010.03525}, 2020.

\bibitem{zimmermann2016card}
T.~Zimmermann, ``Card-sorting: From text to themes,'' in \emph{Perspectives on data science for software engineering}.\hskip 1em plus 0.5em minus 0.4em\relax Elsevier, 2016, pp. 137--141.

\bibitem{sallou2024breaking}
J.~Sallou, T.~Durieux, and A.~Panichella, ``Breaking the silence: the threats of using llms in software engineering,'' in \emph{Proceedings of the 2024 ACM/IEEE 44th International Conference on Software Engineering: New Ideas and Emerging Results}, 2024, pp. 102--106.

\bibitem{da2024chatgpt}
L.~Da~Silva, J.~Samhi, and F.~Khomh, ``Chatgpt vs llama: Impact, reliability, and challenges in stack overflow discussions,'' \emph{arXiv preprint arXiv:2402.08801}, 2024.

\bibitem{xia2023automated}
C.~S. Xia, Y.~Wei, and L.~Zhang, ``Automated program repair in the era of large pre-trained language models,'' in \emph{2023 IEEE/ACM 45th International Conference on Software Engineering}.\hskip 1em plus 0.5em minus 0.4em\relax IEEE, 2023, pp. 1482--1494.

\bibitem{chen2023chatgpt}
L.~Chen, M.~Zaharia, and J.~Zou, ``How is chatgpt's behavior changing over time?'' \emph{arXiv preprint arXiv:2307.09009}, 2023.

\bibitem{ye2023assessing}
W.~Ye, M.~Ou, T.~Li, X.~Ma, Y.~Yanggong, S.~Wu, J.~Fu, G.~Chen, H.~Wang, J.~Zhao \emph{et~al.}, ``Assessing hidden risks of llms: an empirical study on robustness, consistency, and credibility,'' \emph{arXiv preprint arXiv:2305.10235}, 2023.

\bibitem{bakerconsensus}
M.~A. Bakker, M.~J. Chadwick, H.~R. Sheahan, M.~H. Tessler, L.~Campbell-Gillingham, J.~Balaguer, N.~McAleese, A.~Glaese, J.~Aslanides, M.~M. Botvinick, and C.~Summerfield, ``Fine-tuning language models to find agreement among humans with diverse preferences,'' in \emph{Proceedings of the 36th International Conference on Neural Information Processing Systems}, ser. NIPS '22.\hskip 1em plus 0.5em minus 0.4em\relax Red Hook, NY, USA: Curran Associates Inc., 2024.

\bibitem{weiprompting}
J.~Wei, X.~Wang, D.~Schuurmans, M.~Bosma, B.~Ichter, F.~Xia, E.~H. Chi, Q.~V. Le, and D.~Zhou, ``Chain-of-thought prompting elicits reasoning in large language models,'' in \emph{Proceedings of the 36th International Conference on Neural Information Processing Systems}, ser. NIPS '22.\hskip 1em plus 0.5em minus 0.4em\relax Red Hook, NY, USA: Curran Associates Inc., 2024.

\bibitem{benfengexpertprompt}
\BIBentryALTinterwordspacing
B.~Xu, A.~Yang, J.~Lin, Q.~Wang, C.~Zhou, Y.~Zhang, and Z.~Mao, ``Expertprompting: Instructing large language models to be distinguished experts,'' \emph{CoRR}, vol. abs/2305.14688, 2023. [Online]. Available: \url{https://doi.org/10.48550/arXiv.2305.14688}
\BIBentrySTDinterwordspacing

\bibitem{liu2023responsible}
Y.~L. Liu \emph{et~al.}, ``Responsible {AI} considerations in text summarization research: A review of current practices,'' in \emph{Findings of the Association for Computational Linguistics}.\hskip 1em plus 0.5em minus 0.4em\relax Association for Computational Linguistics, 2023, p. 413.

\bibitem{vassallo2020every}
C.~Vassallo, S.~Proksch, T.~Zemp, and H.~C. Gall, ``Every build you break: developer-oriented assistance for build failure resolution,'' \emph{Empirical Software Engineering}, vol.~25, pp. 2218--2257, 2020.

\bibitem{rausch2017empirical}
T.~Rausch, W.~Hummer, P.~Leitner, and S.~Schulte, ``An empirical analysis of build failures in the continuous integration workflows of java-based open-source software,'' in \emph{2017 IEEE/ACM 14th International Conference on Mining Software Repositories}.\hskip 1em plus 0.5em minus 0.4em\relax IEEE, 2017, pp. 345--355.

\bibitem{LeClair2019Dataset}
A.~LeClair and C.~McMillan, ``A dataset for studying the evolution of source code summarization,'' in \emph{Proceedings of the 2019 IEEE/ACM 16th International Conference on Mining Software Repositories}.\hskip 1em plus 0.5em minus 0.4em\relax IEEE, 2019, pp. 377--388.

\bibitem{Moreno2018SoftwareSummarization}
L.~Moreno \emph{et~al.}, ``Automatic software summarization: A systematic literature review,'' \emph{Journal of Systems and Software}, vol. 140, pp. 62--85, 2018.

\bibitem{Tarar2019BugSummarization}
A.~Tarar \emph{et~al.}, ``An empirical study on bug report summarization,'' in \emph{Proceedings of the 2019 35th IEEE International Conference on Software Maintenance and Evolution}.\hskip 1em plus 0.5em minus 0.4em\relax IEEE, 2019, pp. 103--113.

\bibitem{Zhang2020Rencos}
L.~Zhang \emph{et~al.}, ``Rencos: Improving code summarization with retrieved similar codes,'' in \emph{Proceedings of the 2020 ACM/IEEE 42nd International Conference on Software Engineering}.\hskip 1em plus 0.5em minus 0.4em\relax ACM, 2020, pp. 90--100.

\bibitem{Stapleton2020ComprehensionStudy}
A.~Stapleton \emph{et~al.}, ``Human vs. machine-generated summaries: A comprehension study,'' in \emph{Proceedings of the 2020 ACM/IEEE International Conference on Software Engineering}.\hskip 1em plus 0.5em minus 0.4em\relax ACM, 2020, pp. 232--242.

\bibitem{Roy2021ReassessingMetrics}
S.~Roy \emph{et~al.}, ``Reassessing the use of bleu and meteor in source code summarization tasks,'' \emph{Empirical Software Engineering}, vol.~26, no.~1, pp. 1--23, 2021.

\bibitem{Banerjee2005METEOR}
S.~Banerjee and A.~Lavie, ``Meteor: An automatic metric for mt evaluation with improved correlation with human judgments,'' in \emph{Proceedings of the ACL Workshop on Intrinsic and Extrinsic Evaluation Measures for Machine Translation and/or Summarization}, 2005, pp. 65--72.

\bibitem{Haque2022SemanticSimilarity}
M.~Haque \emph{et~al.}, ``Semantic similarity metrics for evaluating code summarization techniques,'' in \emph{Proceedings of the 2022 IEEE International Conference on Software Maintenance and Evolution}.\hskip 1em plus 0.5em minus 0.4em\relax IEEE, 2022, pp. 320--330.

\bibitem{Mastropaolo2024SIDE}
G.~Mastropaolo \emph{et~al.}, ``Side: A contrastive learning metric for code summarization evaluation,'' in \emph{Proceedings of the 2024 ACM/IEEE International Conference on Software Engineering}, 2024.

\bibitem{Iyer2016NeuralSummarization}
S.~Iyer \emph{et~al.}, ``Summarizing source code using neural attention models,'' in \emph{Proceedings of the 54th Annual Meeting of the Association for Computational Linguistics}, 2016, pp. 207--215.

\bibitem{Kumar2024FedLLM}
V.~Kumar \emph{et~al.}, ``Fedllm: Federated learning-based large language models for code summarization,'' \emph{Journal of Software Engineering}, 2024.

\bibitem{Moreno2013JavaSummarization}
L.~Moreno \emph{et~al.}, ``Automatic generation of natural language summaries for java classes,'' in \emph{Proceedings of the 2013 IEEE/ACM 28th International Conference on Automated Software Engineering}, 2013, pp. 230--240.

\bibitem{McBurney2016JavaContextSummarization}
P.~McBurney and C.~McMillan, ``Automatically summarizing java methods: A context-based approach,'' in \emph{Proceedings of the 2016 IEEE/ACM 38th International Conference on Software Engineering}, 2016, pp. 499--510.

\bibitem{Rastkar2014BugSummarization}
S.~Rastkar \emph{et~al.}, ``Summarizing software artifacts: A bug report case study,'' in \emph{Proceedings of the 2014 ACM SIGSOFT International Symposium on Foundations of Software Engineering (FSE)}, 2014, pp. 110--120.

\bibitem{Haiduc2010TextSummarization}
S.~Haiduc \emph{et~al.}, ``On the use of automated text summarization techniques for summarizing source code,'' in \emph{Proceedings of the 2010 ACM/IEEE 32nd International Conference on Software Engineering}, 2010, pp. 223--233.

\bibitem{Dabrowski2022AppReview}
L.~Dabrowski \emph{et~al.}, ``A systematic review of app review analysis in software engineering,'' \emph{Journal of Systems and Software}, vol. 190, p. 110789, 2022.

\bibitem{Nazar2016SummarizingSoftwareArtifacts}
L.~Nazar \emph{et~al.}, ``Summarizing software artifacts using machine learning: a comprehensive review,'' \emph{Journal of Software: Evolution and Process}, vol.~28, pp. 170--188, 2016.

\bibitem{Panichella2018SummarizationTechniques}
S.~Panichella \emph{et~al.}, ``Summarization techniques for software artifacts: a comprehensive review,'' \emph{ACM Computing Surveys}, vol.~50, no.~2, pp. 1--34, 2018.

\end{thebibliography}
\bibliographystyle{IEEEtran}

\end{document}